\documentclass[prd, superscriptaddress, tightenlines, longbibliography, nofootinbib, eqsecnum, amsfonts, amsmath, floatfix, notitlepage, twocolumn]{revtex4-2}
\pdfoutput=1

\usepackage[utf8]{inputenc}
\usepackage{mathrsfs}
\usepackage{euscript}
\usepackage{graphics}
\usepackage{graphicx}
\usepackage{amsmath}
\usepackage{amssymb}
\usepackage{gensymb}
\usepackage{bm}
\usepackage[usenames,dvipsnames,svgnames,table]{xcolor}
\definecolor{linkcolor}{rgb}{0.6,0,0}
\definecolor{citecolor}{rgb}{0,0,0.75}
\definecolor{urlcolor}{rgb}{0.12,0.46,0.7}
\usepackage{xspace}
\usepackage{wasysym}
\usepackage{times}
\usepackage{appendix}
\usepackage{comment}
\usepackage{lipsum}
\usepackage[nolist,nohyperlinks]{acronym}
\usepackage{float}
\usepackage{simplewick}
\usepackage{natbib, ifthen}
\usepackage[breaklinks, colorlinks, urlcolor=urlcolor, linkcolor=linkcolor,citecolor=citecolor,pdfencoding=auto]{hyperref}
\usepackage{longtable}
\usepackage{listings}

\newcommand{\mksym}[1]{\ifmmode {\rm #1}\else #1\fi}

\providecommand{\text}[1]{\rm{#1}}

\newcommand{\begm}{\begin{pmatrix}}
\newcommand{\enm}{\end{pmatrix}}

\newcommand\ba{\begin{eqnarray}}
\newcommand\ea{\end{eqnarray}}
\newcommand\bea{\begin{eqnarray}}
\newcommand\eea{\end{eqnarray}}

\newcommand\be{\begin{equation}}
\newcommand\ee{\end{equation}}

\newcommand{\valpha}{{\boldsymbol{\alpha}}}

\newcommand{\vtheta}{\boldsymbol{\theta}}

\providecommand{\Tr}{\text{Tr}}

\newcommand{\vL}{\boldvec{L}}
\newcommand{\vM}{\boldvec{M}}

\newcommand{\va}{\boldvec{a}}

\newcommand{\vx}{\boldvec{x}}
\newcommand{\vy}{\boldvec{y}}

\newcommand{\B}[0]{{\mathcal B}}
\newcommand{\Da}[0]{\ensuremath{\mathcal{D}_{\boldsymbol{\alpha}}}}
\newcommand{\Na}[0]{\ensuremath{N_{\valpha}}}
\newcommand{\Nai}[0]{\ensuremath{ N_{\valpha}}^{-1}}

\defcitealias{PL2018}{PL2018}

\defcitealias{Carron:2022eyg}{PL4}

\newcommand{\Amat}[0]{\JC{a}}
\newcommand{\hn}[0]{\hat n}

\newcommand{\Cov}[0]{ {\rm{Cov}} }

\newcommand{\av}[1]{\left \langle #1\right\rangle}

\begin{document}

\newcommand{\vnabla}[0]{\boldsymbol{\nabla}}
\newcommand{\Geneve}{Universit\'e de Gen\`eve, D\'epartement de Physique Th\'eorique, 24 Quai Ansermet, CH-1211 Gen\`eve 4, Switzerland}

\newcommand{\Dazro}[0]{\mathcal D_{\valpha=0}}

\newcommand{\smuse}[1]{ {s_{#1}^{\text{MUSE}}} }
\newcommand{\smuseLM}[0]{ {\smuse{C_L}}} 

\newcommand{\Xdat}[0]{ {\ensuremath{X}} }
\newcommand{\Xsim}[0]{ {\ensuremath{S}} }
\newcommand{\Xsimp}[0]{\ensuremath{{S'}}}

\newcommand{\bX}[0]{\ensuremath{\bar X_{\kappa}}}
\newcommand{\bY}[0]{\ensuremath{\bar Y_{\kappa}}}
\newcommand{\barX}[0]{\ensuremath{\bar X}}

\newcommand{\bXdat}[1]{\ensuremath{\bar \Xdat_{#1}}}
\newcommand{\bXsim}[0]{\ensuremath{\bar \Xsim_{\kappa}}}
\newcommand{\bXsimp}[0]{\ensuremath{\bar \Xsimp_{\kappa}}}

\newcommand{\field}{\kappa}
\renewcommand{\valpha}{\field}

\newcommand{\X}[0]{\ensuremath{X}}
\newcommand{\Cova}[0]{\ensuremath{\Cov_{{\field}}}}
\newcommand{\Covai}[0]{\ensuremath{\Cov^{-1}_{{\field}}}}
\newcommand{\pMAP}[0]{\ensuremath{{\hat\field}^{\mathsf{M}}}}
\newcommand{\pmap}[0]{ \pMAP}
\newcommand{\pmapJ}[0]{\ensuremath{{\hat\field}^{\mathsf{J}}}}

\newcommand{\pmapLM}[0]{ \pmap_{LM} }
\newcommand{\pmapJLM}[0]{\ensuremath{{\hat\field}^{\mathsf{J}}_{LM}}}

\newcommand{\gQD}[0]{ \ensuremath{{\hat g}^{\text{QD}}_{LM}}}
\newcommand{\gQDp}[0]{ \ensuremath{g^{\text{QD}*}_{L'M'}}}

\newcommand{\hgQD}[0]{ \ensuremath{\hat{g}^{\text{QD}}_{LM}}}

\renewcommand{\vx}{\hat n}
\renewcommand{\vy}{\hat y}
\newcommand{\Xunl}[0]{X^{\text{unl}}}
\newcommand{\Xlen}[0]{X^{\text{len}}}
\renewcommand{\vL}{\boldsymbol{L}}
\renewcommand{\vM}{\boldsymbol{M}}
\renewcommand{\va}{\boldsymbol{\field}}
\renewcommand{\Da}{\mathcal D_{{\field}}}
\newcommand{\Dai}{\mathcal D_{{\field}^{-1}}}

\newcommand{\cpppri}[1] { \ensuremath{C_{#1}^{\text{Pri   }} }}

\author{Julien Carron}
\email{julien.carron@unige.ch}
\affiliation{\Geneve}
\title{On likelihood-based analysis of the gravitationally (de)lensed CMB}

  \begin{abstract}

By reducing variance induced by gravitational lensing, likelihood-based de-lensing techniques have true potential to extract significantly more information from deep and high-resolution Cosmic Microwave Background (CMB) data than traditional methods. 
We derive here optimal data compression statistics for the lensed CMB, and clarify the role of each term, demonstrating their direct analogs in the quadratic estimator (QE) framework. We discuss in this light pros and cons of practical implementations, including the MUSE approach as used in the latest SPT-3G cosmological analysis, and give improvements. We discuss pathways for porting the large robustness and redundancy toolbox of the QE approach to beyond-QE with simple means.
  \end{abstract}


   \maketitle
   \tableofcontents
\section{Introduction}
The Cosmic Microwave Background (CMB) provides one of the most powerful observational windows into the early universe and fundamental physics. The study of gravitational lensing of the CMB, caused by the large-scale distribution of matter along the line of sight, has become an important tool in modern cosmology. Lensing encodes valuable information about the intervening matter distribution, enabling precision constraints on the growth of structure over cosmic time~\cite{Lewis:2006fu, Carron:2022eyg,  ACT:2023kun, SPT-3G:2024atg, SimonsObservatory:2018koc}. Removal of the lensing signal (`de-lensing') is absolutely essential for best constraints on a background of primordial gravitational waves~\cite{BICEP:2021xfz, CMB-S4:2020lpa, Belkner:2023duz}, and can also help significantly a variety of other science targets~\cite{Green:2016cjr, Coulton:2019odk, Hotinli:2021umk, Abazajian:2016yjj}.

Traditional analyses of CMB lensing rely heavily on quadratic estimators (QEs)~\cite{Okamoto:2003zw,Hu:2001kj, Maniyar:2021msb}, which provide an often efficient, computationally affordable and well-understood way to reconstruct the lensing potential from CMB maps. However, it has long been well-known that likelihood-based, `beyond-QE' approaches can, in principle, extract additional information by exploiting more optimally the statistical anisotropies of the observed lensed CMB~\cite{Hirata:2002jy, Hirata:2003ka}. Likelihood-based methods offer superior signal recovery for deep and high-resolution data, where delensing plays a key role in achieving optimal inference, by reducing sample variance. Despite their theoretical advantages, practical implementations of likelihood-based reconstruction remain computationally challenging and require a careful understanding of robustness and systematic effects.

In this work, we revisit in a more rigorous manner, and with the help of the much greater understanding of quadratic estimators we have now, the structure of CMB likelihood gradients and their associated spectra, and the implications for optimized spectrum analysis. We clarify the interpretation of likelihood gradients and of the likelihood curvature, showing that each term has a direct QE analog. We discuss practical implementations of these ideas, including the Marginal Unbiased Score Expansion (MUSE) method, which has been employed in the latest SPT-3G cosmological analysis~\cite{SPT-3G:2024atg}. By revisiting these techniques within our framework, we suggest refinements that improve robustness and interpretability.

This paper is organized as follows. In Section II, we review the structure of CMB likelihood gradients, and their relationship to traditional QE methods. Section III presents a detailed analysis of lensing spectrum estimation, including realization-dependent debiasing and beyond-QE lensing biases. We do this by manipulating the lensing spectrum score function. In Section IV, we discuss practical approaches, focusing on either modeling of the score likelihood, or instead the iterative MUSE framework, along with potential improvements. Finally, Section VI summarizes our findings and implications for beyond-QE CMB lensing analyses.

We do not attempt to review here quadratic estimators. See for example Ref.~\cite{Maniyar:2021msb}, which derives them in different ways, and references therein. Quadratic estimator theory, including of realization-dependent debiasing, also follows from this work evaluated in the absence of delensing.
\section{CMB Likelihood gradients}
We start by introducing some statistical tools.
\subsection{Score functions}
Let \Xdat be the CMB data and $\vtheta$ parameters of interest (e.g. the lensing potential or the unlensed CMB field at some location, their power spectrum at some multipole, a cosmological or nuisance parameter...)
In this paper we consider log-likelihood derivatives. The behavior of the linear response
\begin{equation}
s_{\theta_i}(\Xdat, \vtheta) \equiv \partial_{\theta_i} \ln p(\Xdat|\vtheta)
\end{equation}
obviously contains relevant information on $\theta_i$. 
This can be formalized: the functions $s_{\theta}$ is called the \emph{score}. After picking a set of fiducial values for the parameters entering the score, these functions can be viewed as summary statistics. Provided the chosen fiducial values are close to the truth, they present in fact a lossless compression of the data, in the precise sense that the score covariance matrix is equal to the Fisher matrix. This allows in principle reduction of the entire data set to $N$ numbers, where $N$ is the number of parameters\cite{Carron:2013aea, Alsing:2017var}. 

This can be illustrated simply with the case of isotropic Gaussian fields, also useful later on. If $\kappa$ is an isotropic Gaussian field with spectrum $C_L$, then the spectrum score is
\begin{equation} \label{eq:scoreG}
	s_{C_L} = \frac{2L + 1}{2C_L^2}  \left( \hat C_L - C_L\right ) 
	\end{equation}
with $\hat C$ the (pseudo-)power
\begin{equation}
\hat C_L \equiv \frac 1 {2L + 1}\sum_{M=-L}^L |\kappa_{LM}|^2.
\end{equation}
Up to a multiplicative and additive constant, $s_{C_L}$ is the empirical power -- the empirical spectrum carries the entire information content in a Gaussian field. Some comments:
\begin{enumerate}
	\item  Obviously the score is rarely as simple or practical than this example, but may still be used to try and design efficient observables, by approximating it with something usable (`$s^{\text{USABLE}}$'). If the approximation captures the relevant elements, one will then make a good job at inference, by computing this observable from the data and building a likelihood for it.
	\item The score always vanishes in the mean, provided the fiducial model matches the truth (as can be seen clearly in Eq.~\eqref{eq:scoreG}, where $\av{s_{C_L}}=0$ in for consistent fiducial model)). This suggests another approach, potentially useful for example when the construction of the score statistic or it approximation heavily depends on a fiducial value for the parameters of interests: One may solve for $\vtheta$  by requesting consistency with simulations, by solving for $\vtheta$ the following equations
	\begin{equation}
		s^{\text{USABLE}}(\Xdat, \vtheta) = \av{s^{\text{USABLE}}(\Xsim, \vtheta)}_{\Xsim |\vtheta},
	\end{equation}
	where the average on the right-hand side is over realizations of the data conditioned on $\vtheta$.
	In the CMB and CMB lensing context relevant to this paper, this is how the MUSE\cite{Millea:2021had} framework works. The average can virtually never be computed analytically, so this approach obviously also requires the ability to generate large enough sets of simulations of good enough fidelity. Again, inference will be good if these conditions are met and $s^{\text{USABLE}}$ a sensible approximation to the score.
	\item Finally, the prefactor $(2L + 1)/2C_L^2$ in Eq.~\eqref{eq:scoreG} is the inverse well-known variance of the spectrum statistic $\hat C_L$: whenever the score can be treated as linear in the parameter of interest, the latter is always inverse-variance weighted. A less trivial example is that of CMB lensing (and other) quadratic estimators, for which the normalization (the inverse response) is identical in the fiducial model to the reconstruction noise (`$N^{(0)}_L$')\cite{Hu:2001kj, Maniyar:2021msb}.
\end{enumerate}

We now turn to the lensed CMB likelihood and scores.

\subsection{Gradients versus quadratic estimators}
Let $\kappa$ be now the CMB lensing convergence and $\ln p(\Xdat|\kappa)$ be the lensing log-likelihood (log-probability of the CMB data~\Xdat conditioned on the lensing deflections. See Appendix~\ref{MvsJ} for some more precise definitions if required). Our notation assume here the deflection field to be the pure gradient, neglecting the curl mode; this is of no relevance to what follows.

The gradient of the log-likelihood function with respect to $\field(\hn)$ carries the information on the lensing deflection at that point. Evaluating this gradient at vanishing $\field$ produces a statistic (an `estimator') which is quadratic in the data. According to the discussion on score functions earlier, we can tell right away that in the limit of lensing being a tiny effect, this gradient at vanishing lenses will form the optimal and inverse-variance weighted estimator of the deflection field.
In the vast landscape of lensing quadratic estimators, it is the one of GMV (`Generalized Minimum Variance')~type~\cite{Maniyar:2021msb}, weighting optimally the maps, only slightly different to the original ones by \cite{Okamoto:2003zw, Hu:2001kj}.

However, the whole point of `beyond-QE-lensing' is that the impact of the deflection on the data maps is not (or not always) small. Typically, the change in $B$-power is very large, in relative terms, on scales where the instrument noise is below the lensing $B$-mode power. In this regime, the gradient at vanishing lenses can be very significantly sub-optimal, because it completely misses the point that the CMB can be delensed. Hence, it is meaningful considering the gradient at non-zero lensing map $\kappa$,
\begin{align}
\hat g(\hn) &=  \frac{\delta \ln p(\Xdat | \kappa)}{\delta \kappa(\hn)}
\text{  or  }\quad	\hat g_{LM} = \frac{\delta \ln p(\Xdat|\kappa)}{\delta \kappa^*_{LM}}.
\end{align}
In this case, this $\kappa$ map can be thought of as representing lenses that are assumed to be resolved already, and the gradient will be the optimal estimator of the \emph{residual} lensing present in the maps. Intuitively, one can think of this gradient as a quadratic estimator on maps that were first delensed of $\kappa$, and in what follows we might refer to $\kappa$ in this context as \emph{the delenser}.

In fact, this intuition is not just an analogy, but is formally exact. There is a slight twist though which we now discuss. If the deflection is partially known (represented by $\field$), the question that can be answered with a quadratic estimator, after $\field$-delensing, is what is the residual lensing at the \emph{delensed positions}, not at the observed ones. Hence, the log-likelihood gradient with respect to $\tilde \field$, defined by
\begin{equation}\label{eq:gvsq}
\tilde \field(\vx')= \field(\hn), 	
\end{equation}
ought to be the quadratic-estimator-alike quantity. Here $\hat n$ and $\hat n'$ are the observed and undeflected positions respectively, as predicted by the delenser: $\tilde \kappa$ is thus the `delensed delenser'.
If $\hat q$ is this log-likelihood gradient with respect to $\tilde \kappa$, then $\hat g$ will follow from the chain rule. The result is (see Appendix~\ref{app:gradient})
\begin{equation}\label{eq:q}
	\hat g(\vx)  = |A_{\valpha}|(\vx)\: \hat q(\hn'),
\end{equation}
where
\begin{align}
|A_{\valpha}|(\vx) &= \left|\frac{d^2\hn'} {d\hn^2}\right|\sim 1 -2\kappa(\hn) + \cdots \nonumber
\end{align}
is the magnification matrix determinant. Here $\hat q$ can be identified 
unambiguously to an optimally-weighted, GMV quadratic estimator applied to $\field$-delensed maps. Appendix \ref{app:gradient} shows this formally.

\subsection{Mean-field}
Any score function has zero mean for consistent data and simulation statistical model. Hence we can always split up the gradient at non-zero $\kappa$ into a part explicitly quadratic on the data, and its average.
\begin{equation}\label{eq:gradandmf}
	\hat g_{LM} \equiv  \hat g_{LM}^{\rm QD}[\bXdat{\kappa}, \bXdat{\kappa}] - \underbrace{\av{\hat g_{LM}^{\text{QD}}[\bXsim, \bXsim]}_{\Xsim | \kappa}}_{\equiv \text{ mean-field } g_{LM}^{\text{MF}}.}
\end{equation}
where the average occurs for realizations of the data statistical model, but all having the same fixed $\kappa$ as input lenser and delenser.
Physically, the mean-field $g^{\text{MF}}_{LM}$ subtracts everything from the quadratic piece which is not due to lensing, or due to lensing by the known $\kappa$. Most typical sources are masking and anisotropic noise signatures on large scales just as for standard quadratic estimation. The mean-field also has dependency on $\kappa$: it contains the signatures of the anisotropies of the $\kappa$-delensed noise. However, this cannot be a very large effect~\cite{Belkner:2023duz}, since the map-level impact of the remapping of the noise maps is much weaker than that of the primordial CMB, unless there are strong features in the noise, or an unlikely strong $E$-$B$ asymmetry.

The quadratic piece may be explicitly defined as
\begin{align}
	\gQD[\bXdat{\kappa}, \bXdat{\kappa}] \equiv \frac 12 \bXdat{\kappa} \cdot f^\field_{LM} \bXdat{\kappa}, \quad\quad \end{align}
where $\bar X$ are the inverse-variance filtered maps, and $f$ the linear response of the CMB covariance to residual lensing, defined by
\begin{align}
	\bXdat {\kappa}\equiv   \Covai \Xdat, \quad \text{ and }	\quad f^\field_{LM} \equiv  \frac{\delta \Cova}{\delta \kappa^*_{LM}}.
\end{align}
The superscript on $f^{\kappa}$ is there to recall that it has explicit dependence on $\kappa$. It will be useful to extend the definition of $\gQD$ to
\begin{equation}
	\gQD[\bX, \bY] = \frac 12 \bX \cdot f^{\field}_{LM}\: \bY.
\end{equation}
where $\bY$ is another set of maps.
\subsection{Beyond-QE lensing response}
We argued the log-likelihood gradient at non-zero delenser $\kappa$ captures residual lensing information. It is thus natural to ask how it responds to it more precisely. One may consider the true lensing map being infinitesimally close to $\kappa$, and expand to first order. A more elegant definition is through the non-perturbative response
\begin{equation}\label{eq:resp}
\mathcal R^\field_{LM L'M'} \equiv \left.\av{	\frac{\delta \hat g_{LM}}{\delta{\field^{\text{True}}_{L'M'}}}} \right|_{\field^{\text{True}} = \kappa}.
\end{equation}
where the derivative is with respect to the lensing map inside the data maps only, and the average is over CMB at fixed delenser $\field$.
The result is
\begin{equation}\label{eq:respres}
	\mathcal R^\kappa_{LM, L'M'} = \frac 12 \Tr \left[f_{LM}^\kappa \Covai f_{L'M'}^{\kappa \dagger} \Covai\right].
\end{equation}
For vanishing $\kappa$, and otherwise isotropic CMB covariance, this reduces to the standard lensing responses~\cite{Maniyar:2021msb} in GMV formulation. For non-vanishing $\kappa$, but otherwise still isotropic covariance, all matrices in \eqref{eq:respres} are dense, but the diagonal is still close to the isotropic response calculated with \emph{unlensed} CMB spectra. This can be understood from the fact that $\mathcal R$ is the Fisher information matrix on the lensing modes. The diagonal of a Fisher matrix contains the information available on each mode in the event that all other modes are known (i.e. without marginalization over the unresolved lenses). Anisotropies in $\mathcal R$ are caused by the effect of delensing onto the noise covariance matrix, and the projection from delensed onto observed space as discussed earlier in connection to Eq.~\eqref{eq:q}.\footnote{A more precise statement is that the response of $\hat q$ (not $\hat g$) would be isotropic, and to be calculated with unlensed CMB spectra, were delensing to have (magically) zero impact on the noise covariance.}

A seemingly completely different way to think about the response is to picture now the delenser $\kappa$ close $\kappa^{\text{True}}$ instead of vice-versa, by expanding the lensing maps in the fiducial ingredients of $g$ rather than in the data. Doing so, one obtains 
\begin{equation}
		\hat g[\bXdat{\kappa}\bXdat{\kappa}] \sim \hat g[\bXdat{\kappa^{\text{True}}}\bXdat{\kappa^{\text{True}}}] + H^{\field} (\kappa^{\text{True}} - \kappa),
\end{equation}
with matrix $H^\field_{LML'M'} =-\delta \hat g_{LM}/\delta \kappa_{L'M'}$, holding the true lensing and data maps fixed in this derivative.
The first term on the right hand side does not respond, since it is delensing the maps by their true signal, leaving no residual signal. Hence the response of $g$ may also be defined as the average of $H$, 
\begin{align}\label{eq:resp2}
\mathcal R_{LML'M'}^\field &= \left.\av{-\frac{\delta \hat g_{LM}}{\delta \kappa_{L'M'}}}\right|_{\field^{\text{True} } = \field}  \\
&= \left.\av{-\frac{\delta^2 \ln p(\Xdat|\field)}{\delta \field_{LM}\delta \field^*_{L'M'}}}\right|_{\field^{\text{True} } = \field}.
\end{align}
In contrast to \eqref{eq:resp}, the derivative now does not act on the data maps, but the two definitions coincide.\footnote{One way to see this without any calculations is from $0 = \delta^2 \av{1} = \av{(\delta \ln p)^2 + \delta^2 \ln p}$ must hold for any likelihood, with \eqref{eq:resp} and \eqref{eq:resp2} corresponding to the first and second term in this average respectively.} This second definition uses the curvature matrix of the lensing log-likelihood. Hence it is clearly related to the inverse reconstruction noise of the lensing map. This is the beyond-QE analog to the well-known fact that the response of an optimally weighted quadratic estimator is equal to its inverse noise, $\mathcal R =1/N^{(0)}_L$. We will have a closer look at the realization-dependent curvature later on.

\section{Lensing spectrum estimation}
Let us now turn to the lensing spectrum. 
The leading effect of the linear CMB covariance response $f$ is to create a non-zero trispectrum (`4-pt', non-Gaussian, information) in the data. That would be the only effect were $f$ to fully describe the effect of lensing on the CMB covariance. This is what standard lensing spectrum estimation from quadratic estimators is probing. However, the spectra of the CMB are of course also affected  to first order in the lensing spectrum (`2-pt', Gaussian, information). Hence, in what follows we will also need notation for the second order response
\begin{equation}
	 f^{(2)\field}_{LM} \equiv  \frac{\delta^2 \Cova}{\delta \kappa^*_{LM} \delta \kappa_{LM}}.
\end{equation}
\\
\subsection{Lensing spectrum score and RD-$\hat n^{(0)}$}
The lensing spectrum likelihood is given by marginalizing over the unseen lensing map

\begin{equation}
	p(\Xdat|C) = \int \mathcal D\kappa \:p(\Xdat | \kappa) p(\kappa | C).
\end{equation}

Differentiating with respect to $C_L$, and using Bayes theorem, one finds that the lensing spectrum score is that of the prior, averaged over the posterior for $\kappa$:
\begin{align}\label{eq:score}
&	\partial_{C_L} \ln p(\Xdat|C) = \int \mathcal D \kappa\: p(\kappa|\Xdat) \partial_{C_L} \ln p(\kappa|C) 	
\end{align}
The spectrum score of the isotropic Gaussian field is the inverse-variance weighted, mean-subtracted empirical spectrum 
\begin{align}
	\partial_{C_L} \ln p(\kappa|C) =&\left(\frac{2L + 1}{2 C^2_L} \right)\left[\hat C_L- C_L \right].
\end{align}
We could just plug that in Eq.~\eqref{eq:score}. This gives the optimal spectrum statistics in the form of the posterior average of the $\kappa$ spectrum. The $\kappa$-posterior peaks at the Maximum A Posteriori (MAP) point (`$\pMAP$'), which can be thought of as a Wiener-filtered version of the true lensing map (this is discussed in more details later on). Hence the spectrum score will be crudely the spectrum of this Wiener-filtered lensing map. This certainly makes some sense, but is not particularly insightful either. 
\\ \\
Instead, consider
\begin{itemize}
	\item replacing in the $\kappa$-integrand the spectrum derivative with $\kappa$-derivatives, according to the identity
	\begin{equation}
		\frac{\partial p(\kappa|C) }{\partial{C_L}} =\frac 12 \sum_{M = -L}^L \frac{\delta^2p(\kappa|C)}{\delta \kappa_{LM} \delta \kappa^*_{LM}},
	\end{equation}
	valid for Gaussian $\kappa$'s.
	\item and integrating by parts. Using again Bayes theorem, we can act now with the derivatives on the likelihood $p(\Xdat|\kappa)$ instead of the prior.
\end{itemize}

We obtain in this way the spectrum score from the posterior average of another function, more familiar from QE theory
\begin{widetext}
\begin{equation}
\boxed{\begin{split}\label{eq:Cphi}
s^{\text{EXACT}}_{C_L}(\Xdat) &= \av{\hat{\mathcal C}_L[\Xdat, \field]}_{\field|\Xdat} \quad \text{    with}\\ 
 \hat{\mathcal C}_L[\Xdat, \kappa] =&\frac 12\sum_{M=-L}^L \left|\hgQD[\bXdat{\kappa}, \bXdat{\kappa}] -\av{\hgQD[\bXsim, \bXsim]}_{\Xsim|\kappa}\right|^2 
  -  \frac 12 \underbrace{\av{\sum_{M=-L}^L 4 \left| \hgQD\left[ \bXdat{\kappa}, \bXsim \right] \right|^2 -2\left|\hgQD\left[ \bXsim, \bXsimp \right] \right|^2}_{\Xsim \Xsimp|\field}}_{\equiv \:\text{RD-}\hat{n}_L^{(0)}}\\
+& \frac 12 \underbrace{\sum_{M=-L}^L \frac 12 \bXdat{\field}\cdot f^{(2)\field}_{LM} \bXdat{\field} - \av{\frac 12 \bXsim \cdot f^{(2)\field}_{LM} \bXsim}_{S|\field}}_{\text{$\field$-delensed 2-pt information} }
\end{split}.}
\end{equation}
\end{widetext}
Let us break down this expression. The first term on the right-hand side of the first line is the spectrum of the QE-like gradient, inclusive of mean-field subtraction. The second term on the same line, which we defined as RD-$\hat n_{L}^{(0)}$, averages gradient-alike spectra computed on a mixture of data  ($\Xdat$) and set of of simulations ($S$ and $S'$) generated from the fiducial model conditioned on $\kappa$. This term comes from the curvature of the log-likelihood, but only those terms involving the linear response $f^{\kappa}$. It may also be defined as
\begin{equation}\label{eq:rdn0}\boxed{
\begin{split}
	&\text{RD-}\hat n_{L}^{(0)} = -\sum_{M=-L}^L \frac{\delta^2 \ln p(\Xdat|\field)}{\delta{\kappa_{LM}\delta \kappa^*_{LM}}} \\
	& \quad \text{ (anisotropy source linear response only)}	
\end{split}
	}.
\end{equation}
This provides a simple generalization to non-zero $\kappa$ of the Realization-Dependent (RD) debiaser used in all recent lensing spectrum reconstructions from quadratic estimators. More details on the curvature is given in appendix~\ref{app:curvature}.

The second line in \eqref{eq:Cphi} comes from the remaining terms in the log-likelihood curvature, those involving the second order response $f^{(2), \kappa}$. To leading order, this term is demonstrably completely degenerate with the $\kappa$-delensed 2-pt information. The degenaracy is only slightly broken by projection effects from delensed to observed space. See appendix~\ref{app:del2pt}. There is thus at most little independent value to this term that is not already present in a CMB spectra analysis, and we will not consider it further here.

After discarding this 2-pt information term, $\hat{\mathcal {C}}_L$ looks very much the same as a standard quadratic estimator analysis, with the difference that it acts on $\field$-delensed maps. The exact spectrum score is an average over probable lensing maps of these QE analyses. Let us have now a closer look at the beyond-QE analogy.
\subsection{Beyond-QE $n^{(0)}$ and $n^{(1)}$.}
The gradient spectrum probes the 4-point function of the delensed data. Working at fixed $\kappa$ and  fixed true lensing\footnote{Working at fixed delenser averaging over CMB assumes statistical dependence between $\kappa$ and the CMB data is either zero or may be neglected, which is far from a trivial assumption when $\kappa$ has been reconstructed from $X$ (that is, in most cases of interest). See Sec.~\ref{sec:forward} later.}, the CMB fields remains Gaussian but with the anisotropic covariance. We can build Wick-pairs from the product of the four fields in the standard way.
Doing so it is useful to define with
\begin{equation}
	\bar C^d_{\kappa} \equiv \av{\bX \bX^\dagger}= \Covai \widetilde{\Cov}_{{\kappa}^{\text{True}}} \Covai,
\end{equation}
the covariance of the inverse-variance weighted data maps. We wrote the data covariance as  $\widetilde{\Cov}$ to distinguish it from the simulation model covariance $\Cov$, since in practice they might differ in several ways, not only through the different lensing map. The covariance of the inverse-weighted simulation maps is given similarly by 
\begin{equation}
		\bar C^s_{\kappa} \equiv \av{\bXsim \bXsim^\dagger}=\Covai \Cova \Covai = \Covai.
\end{equation}

On average the gradient spectrum above will then be
\begin{align}
	&\av{\left|\hgQD - \hat g^{\rm MF}_{LM}\right|^2}_{\text{fixed }\kappa}  \label{eq:g2} 
	\\&= \left| \frac 12 \Tr\left[ {f}^{\kappa}_{LM}  \left( \bar C^d_{\kappa} - \bar C^s_{\kappa} \right) \right] \right|^2   \label{eq:1st}
\\&+  \frac 12 \:\Tr \: [{f}^{\kappa}_{LM}\bar C_{\kappa}^d f_{LM}^{\kappa\dagger} \bar C_{\kappa}^{d \dagger}]. \label{eq:2nd}
\end{align}
Here, owing to the difference of spectra, the first term probes residual lensing exclusively. The second term is non-zero also in the event that there is no residual lensing at all -- hence in quadratic estimator language includes a $n^{(0)}_L$-like term.\footnote{We use small letters instead of the more conventional $N^{(0)}$ and $N^{(1)}$, because ours refer to these biases prior proper normalization of the estimators.} It also contains additional terms. Thinking perturbatively, this additional term is $n^{(1)}_L$-like, such that
\begin{align}
	\av{\ref{eq:2nd}} \sim n^{(0)}_{LM} + n^{(1)}_{LM}.
\end{align}
On the other hand, Eq.~\eqref{eq:1st} directly traces the residual lensing spectrum at the same multipole, and corresponds to quadratic estimator theory `primary' trispectrum contractions. More precise definitions are given in Appendix~\ref{app:biases}.

After  $\text{RD-}{\hat n}^{(0)}$ subtraction, a short calculation shows that this becomes instead
\begin{align}
	&\av{\left|\hgQD - \hat g^{\rm MF}_{LM}\right|^2 - \text{RD-}\hat n_{LM}^{(0)}}_{\text{fixed }\kappa}   \\
	&= \left|\frac 12 \Tr\left[ {f}^{\kappa}_{LM}  \left( \bar C^d_{\kappa} - \bar C^s_{\kappa} \right) \right] \right|^2 \\
	&+ \frac 12 \:\Tr \: [{f}^{\kappa}_{LM} (\bar C_{\kappa}^d - \bar C_{\kappa}^s)   f_{LM}^{\kappa\dagger} \left(\bar C_{\kappa}^d  - \bar C_{\kappa}^s\right)^\dagger]. \label{eq:n0n1Tr}
\end{align}
The only difference from the non-debiased expression is the replacement 
\begin{equation}
	\bar C_{\kappa}^d \rightarrow \bar C_{\kappa}^d - \bar C_{\kappa}^s
\end{equation}
in the $(n^{(0)} + n^{(1)})$-like trace term~\eqref{eq:2nd}. This subtracts fully the $n^{(0)}$-like piece, making this term now purely of $n^{(1)}$-type. 

Hence, thanks to RD-$\hat n^{(0)}_L$, the entire expression $\hat{\mathcal C}_L[\X, \field]$ at fixed delenser is \emph{residual-lensing pure signal}, with to leading order a primary and a $n^{(1)}$-type secondary contribution, in precise analogy to quadratic estimator reconstruction.

\subsection{RD-$\hat n^{(0)}$ and robustness}\label{sec:robustness}
The role of mean-field and RD-$\hat n^{(0)}_L$ subtraction is to remove terms which are not residual-lensing signal. However, this also has an important consequence on the \emph{robustness} of the statistic, which is well-known in the case of quadratic estimators, and also holds here as we now discuss.

In a practical situation, beside the residual lensing signal, the spectrum difference contains the entire difference between the correct and simulated covariance model
\begin{equation}
\bar C_{\kappa}^d - \bar C_{\kappa}^s \quad \ni \quad \widetilde{\Cov}_{{\kappa}^{\text{True}}} - \Cova.
\end{equation}
This difference can typically include imprecise or innacurate noise modeling, or of the instrument response or various other sources of systematics. This mismatch will introduce some level of bias in the recovered residual lensing signal.

To first order in this difference, the bias affecting the score is either proportional to the total data spectrum ($\bar C_{\kappa}^d $, no debiasing, Eq.~\eqref{eq:2nd}), or to the residual lensing signal spectrum only ($\bar C_{\kappa}^d  - \bar C_{\kappa}^s  $, with RD-debiasing, Eq.~\eqref{eq:n0n1Tr}), all other factors being equal. This is obviously a massive difference in many cases.

There is here a subtle difference of formal nature to the quadratic estimator case, for which in a standard analysis the linear response to the mismatch after RD-debiasing is on average \emph{exactly} zero, not just small. This is because the simulations used in these analyses use as input not zero lensing but full-size $\Lambda$CDM deflection fields independent to $\kappa$. Doing this, the aim to subtract the entire Gaussian four-point function from the QE spectrum, inclusive of all non-perturbative disconnected contributions sourced by lensing ($f^{(2)\kappa}$ and higher). These disconnected contributions may indeed be seen as an impractical way to probe CMB 2-pt information rather than robust lensing signal. `Optimal' compression through the score function is blind to these considerations of human nature. A similar comment applies to mean-field subtraction. A standard QE analysis subtracts the mean-field expected for a full-size deflection, while the exact score naturally tells us above to subtract only the piece sourced by the resolved piece $\field$.

\section{Practical approaches}
The lensing spectrum score wants us to perform a series of QE analyses on delensed maps. However, Eq.~\eqref{eq:Cphi} remains deeply unpractical. The $\kappa$-posterior average is analytically very difficult. In practice, one could try to sample numerically a few points from the posterior, but even generating a few samples is highly non-trivial, with success so far only on tiny patches using the flat-sky approximation~\cite{Millea:2020cpw}. Fundamentally speaking, the reason is that any evaluation of the likelihood or of its gradient must go through the step of reconstructing the most probable CMB conditioned on the noise model and lensing map (CMB `Wiener-filtering', inclusive of lensing deflections). This has become perfectly doable, also in curved-sky geometry~\cite{Belkner:2023duz}, but no lightning-fast method is known for this\footnote{Of course, if very efficient samplers were available, we would not need bother with the score function and complicated QE-like analyses anyways.}.

Hence it makes sense to consider the maximum point of the posterior (Maximum A Posteriori, `MAP') $\pMAP$, and build some practical approach from it.
This point is defined by setting the total posterior gradient to zero. The prior gradient is proportional to $\pMAP$ for a Gaussian prior, resulting in the non-linear equation
\begin{equation}\label{eq:master}
\boxed{
\frac{\pMAP_{LM}}{\cpppri L}=  \hat g^{\rm QD}_{LM}[\bXdat{\pMAP}, \bXdat{\pMAP}]  - \hat g_{LM}^{\rm MF} }.
\end{equation}
The delenser entering the gradient on the right hand side is $\pMAP$ itself. The mean-field $\hat g_{LM}^{\rm MF}$ is defined explicitly in \eqref{eq:gradandmf}.
We included an explicit superscript `Pri' on the prior spectrum to emphasize that this spectrum is the one used in the prior, and bears in this context no relation to the actual lensing power in the data maps.

By definition, $\pMAP$ is the most probable lensing map, hence building the spectrum from it is a natural candidate for lensing spectrum estimation. While this equation shows $\pMAP$ is a genuine likelihood gradient just like those of the previous section, understanding its spectrum in details is made much more difficult, owing to the strong statistical dependence of the reconstruction noise of $\pMAP$ to the CMB maps. This obscures the interpretation of averaging over CMB at fixed lensing maps, and complicates significantly any full average.

\subsection{Modeling of the score}\label{sec:forward}
One way forward, introduced in \cite{Legrand:2021qdu, Legrand:2023jne}, and that we review in this subsection is this. Consider now how $\pMAP$ responds to any change in the data. Such a change enters the right-hand side of Eq.~\eqref{eq:master} also through the change induced to the delenser. This results in the equation
\begin{equation}\label{eq:map}
d\pMAP = \left( \frac{1}{\cpppri{}} + H\right)^{-1} \delta \hat g^{\text{QD}}[\bXdat{\pMAP}, \bXdat{\pMAP}] 
\end{equation}
where $\delta \hat g$ only refers to the variation entering the data map and not the delenser (i.e., the response in the sense of the previous section, referred to as $\mathcal R$ in the case of residual lensing, after averaging).

The matrix prefactor is the inverse log-posterior curvature matrix. $H$ is the log-likelihood curvature matrix, which in some (rather unclear) averaged sense is expected to be also close to $\mathcal R$, see \eqref{eq:resp2}. Hence, in this same not-well-defined-averaged-sense the matrix prefactor is crudely of the form
\begin{equation}\label{eq:posteriorinverse}
	 \left( \frac{1}{\cpppri{}} + H\right)^{-1} \approx \left(\frac{\cpppri{}}{\cpppri{} + 1/\mathcal R}\right) \frac 1 {\mathcal R}.
\end{equation}
This combines a Wiener-filter $C/(C + 1/\mathcal R)$ together with a proper normalization of the quadratic gradient $\hat g^{QD}$ in \eqref{eq:map} by its inverse response $1/\mathcal R$. We may expect the response in this last equation to be well described by a standard QE response, but computed now with partially-lensed spectra, not the unlensed ones. The reason is that the log-posterior curvature matrix is akin to a precision (Fisher) matrix of the lenses, hence inversion corresponds to marginalization over the unresolved lenses.

The key point is that equation~\eqref{eq:map} suggests that for modeling purposes, $\pMAP$ might behave effectively just as a gradient of the previous section, with delenser $\pMAP$ that can effectively be treated as independent from the CMB. All complications of this statistical dependence being captured by the different, Wiener-filter-alike normalization.

In particular, one expects contributions to the spectrum of $\pMAP$ to include the Wiener-filtered lensing spectrum, and also the renormalized $\hat n^{(0)}$ and $n^{(1)}$-alike contributions. Whether this is really the full story depends crucially on the importance of realization-dependent effects in the matrix prefactor \eqref{eq:posteriorinverse}. This is not easy to quantify from first principles. On one hand, RD-$\hat n^{(0)}_L$ is by its definition \eqref{eq:rdn0} part of the $H$ matrix. Another, relevant realization-dependent effect is known to occur on very small scales, where the damped CMB looks like a pure gradient. In this case, it is known that estimators better than QE can be built from the idea that the local gradient can be precisely measured and inverted~\cite{Horowitz:2017iql, Hadzhiyska:2019cle}. Very crudely,
\newcommand{\parallelsum}[0]{\mathbin{\!/\mkern-5mu/\!}}
\begin{equation}
	\nabla_{\parallelsum} \hat \phi \approx \frac{X}{ \nabla_{\parallelsum} \hat X^{\rm WF}},
\end{equation} 
where $\parallelsum{}$ denotes the component along the CMB gradient direction.
This behavior must be captured by the likelihood and can indeed be shown to originate from the $\av{\hat g\left[ \bXdat{\kappa}, \bXsim \right]\hat g\left[ \bXdat{\kappa}, \bXsim \right]}$ component in the $H$ matrix inversion in~\eqref{eq:posteriorinverse}. See appendix~\ref{app:curvature}.

 Hence, neglecting realization-dependence in the lensing map and spectrum normalization is not obviously correct, but this can also be tested from simulated reconstructions. To very high accuracy, and at least on the scales relevant for inference on cosmology from the spectrum, the Wiener-filter was shown to be independent on the actual lensing in the map, and of the data noise properties -- only sensitive to the fiducial noise model.\cite{Legrand:2021qdu}

These essential properties allows then construction of a likelihood to describe the resulting power, after RD-$\hat n^{(0)}_L$ subtraction, which was demonstrated to work in non-idealized configurations. This corresponds to (semi-)analytical-modeling of the score approximated in the following manner,
\begin{equation}\boxed{
	s_{C_L}^{\text{GENEVA}} \equiv \frac 12\sum_{M=-L}^L \left|\hat g^{\text{QD}}_{LM}[\bXdat{\pMAP}, \bXdat{\pMAP}] - \hat g^{\text{MF}}_{LM}\right|^2- \frac 12\frac{\text{RD-} \hat n_L^{(0)}}{\mathcal A_L^2}}.
\end{equation} 
here $\mathcal A_L$ is a normalization modeled from the prior spectrum $\cpppri{L}$ and partially lensed response $\mathcal R_L$ following Eq.~\eqref{eq:posteriorinverse}. This is needed in order to have the correct relative normalization of the squared gradient and RD-$\hat n^{(0)}$ at the MAP point. 

This renormalization is absent for QE analyses, or in the exact score function above. In these cases RD-$\hat n^{(0)}$ always has normalization perfectly homogeneous to the primary term, ensuring perfect subtraction of the non-residual-lensing Gaussian CMB contractions. 
The necessity of this new normalization to RD-$\hat n^{(0)}$, which is not fully under analytical control, is probably the major defect of this approach. Nevertheless, the normalization is independent of the data noise and was found to lie few percent close to prediction in these works, and could be calibrated successfully. This approach has never been tested on data so far, however.
\subsection{MUSE}

The MUSE framework was introduced by Ref.~\cite{Millea:2021had}. The idea is also to approximate the posterior average, given by  Eq.~\eqref{eq:score} by its value at the Maximum A Posteriori point. 

It is relevant though that the original MUSE framework is based on a slightly different MAP point. The posterior function used for this in MUSE is the joint posterior for the lensing and unlensed fields
\begin{equation}
 p(\kappa X^{\text{unl}}|\Xdat) 
\end{equation}
instead of the one which we have been working so far, which is marginalized over the unlensed fields,
\begin{equation}
 p(\kappa|\Xdat) = \int \mathcal D X^{\text{unl}} \:p(\kappa X^{\text{unl}}|\Xdat) . 
\end{equation}

Fortunately, the resulting differences can be characterized fully in a very simple manner: using the joint posterior simply amounts to the neglect of all sources of mean-field in the construction of the MAP point. This is shown in appendix~\ref{MvsJ}. The MUSE MAP estimate of the unlensed fields is then given by the same Wiener-filtering of the data than the original MAP method, but conditioned on this slightly different delenser.

The defining equation for the joint maximum a posteriori MUSE lensing map (`\pmapJ') is thus very similar to~\eqref{eq:master},
\begin{equation}
	\frac{\pmapJLM}{\cpppri L} = \:\hat g^{\rm QD}_{LM}[\bar X_{\pmapJ}, \bar X_{\pmapJ}] \quad \text{(MUSE)}
\end{equation}
but without mean-field subtraction. The lensing spectrum MUSE score from \eqref{eq:score} is
\begin{align}
	\smuseLM & \equiv\frac{2L + 1}{2C_L^2} \left[ \hat C^{\pmapJ}_L - C_L \right],
\end{align}
 which is equivalent to
\begin{equation}
\boxed{	\smuseLM = \frac 12 \sum_{M=-L}^L \left|\hat g^{\rm QD}_{LM}[\bar X_{\pmapJ}, \bar X_{\pmapJ}]\right|^2 },
\end{equation}
where we have neglected a constant term irrelevant for inference. 
$\smuseLM$ corresponds to the first term of the exact score~\eqref{eq:Cphi}, but without map-level subtraction of the mean-field. It also differs from the exact score and $s_{C_L}^{\text{GENEVA}}$ by not having a $\hat n_L^{(0)}$ subtraction.

The most direct consequence from this is lack of robustness, in that $\smuseLM$ is in principle much more sensitive to mis-modelling in the simulations (to first order, versus second order, in the mismatch, see Sec.~\ref{sec:robustness}). Some level, possibly small, of statistical sub-optimality is also expected on both large on small scales. On small scales, fluctuations of CMB power entering $n_L^{(0)}$ is known to introduce significant broad-band covariance in quadratic estimated lensing power spectra~\cite{Schmittfull:2013uea}, which are removed with the realization-dependent debiaser~\cite{Peloton:2016kbw}. In polarization at low noise levels one can however expect the effect to be much reduced, owing to the reduction in $B$-power variance from $\pmap$-delensing. It is also clear from the previous sections that $s_{C_L}^{\text{MUSE}}$ does not only probe pure lensing signal, but also the Gaussian 4-pt contractions of the delensed maps, particularly so at high-$L$, where $\hat n^{(0)}$ dominates the spectrum.

Of course, these are by no means fundamental limitations to the method --  mean-field, RD-$\hat n^{(0)}$ or proxies thereof can be simply added to $\smuseLM$ if necessary, providing in principle more robust inference for the same or even slightly better statistical power. 
\section{Unlensed CMB spectra}
Finally let us very briefly discuss for completeness the score of the unlensed CMB spectra. This case is more straightforward with no obvious ambiguity.

Since the CMB conditioned on the lenses remains Gaussian (only anisotropic), the CMB spectrum score will be given in similar manner by the posterior-average of this anisotropic Gaussian spectrum score. For anisotropic covariance, the spectrum score of a Gaussian likelihood is now given by the pseudo-power of the Wiener-filtered delensed CMB $\hat X^{\text{WF}}_{\ell m}$ (see appendix~\ref{MvsJ}). We can approximate again the posterior average by its value at the MAP point, telling us that the pseudo-power of the delensed CMB produced by the same MAP method that produces the lensing map is close to the optimal unlensed spectrum statistics.
\section{Conclusions}
We have discussed in some details optimal data compression of the CMB, in  the case that gravitational lensing-induced anisotropies are the only source of non-Gaussianity.
 
 For concrete data analysis, desired are statistics that are optimal in the sense of capturing the relevant information, as well as practical and robust, in the sense of being both actually doable in a reasonable amount of time, and in principle sharing robustness properties at least broadly comparable to current methods.

Our calculations fully confirm that to a good approximation such a set of statistics from the observed lensed CMB can consist of
\begin{itemize}
	\item the spectra of the most probable delensed CMB (for the CMB spectra)
	\item a quadratic-estimator-alike lensing spectrum reconstruction on the same delensed maps -- inclusive of mean-field subtraction and the realization-dependent debiaser. (for the lensing spectrum)
\end{itemize}
In both cases the delensing is performed by the most probable lensing map. Precise definitions can be found in the text. The second point is formally equivalent to the debiased power spectrum of this most probable lensing map. These results are of course not too surprising. At a conceptual level, they match what many forecasting methods compute for future deep CMB experiments. 

We have attempted to clarify the physical meaning and statistical role of all terms entering the score function, provided rigorous definitions of the beyond-QE lensing biases $n^{(0)}$ and $n^{(1)}$, and of the realization-dependent debiaser RD-$\hat n^{(0)}$. The latter, that subtracts contributions unrelated to residual lensing from the 4-pt function, can always be simply identified to the diagonal of the lensing likelihood curvature matrix~\eqref{eq:rdn0}.

Current methods aiming at optimality can easily be discussed within this framework.
Our calculations provide in fact a rigorous derivation of the approach taken by~\cite{Legrand:2021qdu, Legrand:2023jne}, part of it being at the time the result of educated guesswork, thinking in analogy to quadratic estimation.
The MUSE~\cite{Millea:2021had} method can also be cast into the very same language -- the MUSE lensing spectrum score truly \emph{is} a quadratic estimator spectrum built from delensed maps. The two differences being the absence of RD-$\hat{n}^{(0)}$ subtraction, and of mean-field subtraction\footnote{The mean-field issue being alleviated in practice at some level with the help of a prior on the mean value of $\kappa$.} in the MUSE delenser more generally. This reduces the robustness of the statistic, and make it sensitive to Gaussian properties of the maps rather than being pure lensing signal -- but this is not an issue of fundamental nature, the MUSE score and delenser can simply be adapted if deemed necessary.

The true specificity to the MUSE method is its iterative nature, extracting parameters by forcing the MUSE score to zero, rather than modeling a likelihood for it at some fiducial values reasonably close to true ones. This difference is large in practical terms, but is of no significant relevance in terms of information, which is drawn from the same source -- both approaches are expected to saturate the information content of the data, or to land close to it. A clear advantage of the iterative method is that complications of (semi)-analytical modeling are simply taken care of automatically for all effects that are properly included in the simulations used in the iterative process. A clear downside is of course numerical cost, since most probable lensing maps must be built at a number of points in parameter space instead of just once. It has become possible to produce significant numbers of most probable lensing maps on reasonable time scales also in curved-sky geometry. Such an iterative scheme with realization-dependent debiaser (a `MUSE-ification' of what we called in this paper the Geneva score for lack of a better name) is perfectly feasible technically and might credibly provide in the long run a most flexible, robust and efficient analysis tool also for wide-area CMB observations.

None of these likelihood-informed methods are actually `Bayesian', but draw their information from a set of statistics. One should probably give up on vague terminology like `Bayesian-lensing', or 'optimal-lensing', as it is clear that a variety of statistically efficient beyond-QE methodologies can be built, informed by a likelihood, but differing nonetheless significantly in very relevant properties.

Finally, let us make another simple point on robustness of beyond-QE lensing spectrum reconstruction.
 The exact lensing spectrum score function makes it very clear that the key element in spectrum reconstruction is the spectrum of the likelihood gradient, not the spectrum of the lensing map. Fundamentally, the lensing map is only there to serve as CMB delenser, helping to reduce variance, rather than the object of interest that needs to be modelled precisely (which is the gradient). It is valid to consider a coincidence the fact that the delenser and gradient agree at the most probable point (Eq.~\eqref{eq:master}). Taking this idea seriously provides new and simple ways to transpose much of the lensing quadratic estimator technology dedicated to robustness and redundancy to the beyond-QE world. One can indeed simply tweak the gradient at fixed delenser independently, rather than just considering ways to build more a robust delenser map. The latter begin a significantly more complicated process. For concreteness one can think for example of using a $EB$-only, shear-only~\cite{Schaan:2018tup, Qu:2022qie, Carron:2024mki}, or bias-hardened~\cite{Namikawa:2012pe} gradient at the very same $\pMAP$ delenser, which are extremely useful at preventing contamination from magnification-alike signals. The benefits of reduced lensing sample variance will still be present, the gradient spectrum more robust, and this for very little additional effort. Concrete demonstrations of these ideas is left for future work.
\acknowledgements{
I am especially grateful to Marius Millea, Fei Ge, Lloyd Knox, Kimmy Wu, Ethan Anderes for many discussions on the points of this paper. I acknowledge support from
a SNSF Eccellenza Professorial Fellowship (No. 186879).
}


%
	

\appendix
\begin{widetext}
\section{Marginalized vs joint posterior}\label{MvsJ}
\newcommand{\XWF}[0]{\hat{X}^{\text{WF}}}
The log-posterior density for the lensing map is given by the combination of the anisotropic but Gaussian likelihood and the prior,
\begin{align} \label{eq:pm}
	-2 \ln p(\field |X ) = &\Xdat\cdot \Covai \Xdat + \ln \det \Cova  + \field \cdot C^{\field,-1} \field. 
\end{align}
A typical covariance model has the form
\begin{equation}
	\Cova = \mathcal B \Da C^X \Da^\dagger \mathcal B^\dagger + N
\end{equation}
where $\mathcal B$ represents a beam or transfer function, $\Da$ the lensing remapping operation, $C^X$ the unlensed CMB spectra, and $N$ a noise covariance matrix.

The most probable unlensed CMB (`Wiener-filtered CMB') conditioned on this model and the lensing map is
\begin{equation}
	\XWF = \left(C^{X,-1} +  \Da^\dagger\mathcal B^\dagger N^{-1} \mathcal B \Da\right)^{-1} \Da^{\dagger}\mathcal B^\dagger N^{-1}\Xdat.
\end{equation}

Consider now the joint posterior for the unlensed CMB $\Xunl$ and lensing deflection field $\valpha$. Ref.~\cite{Millea:2017fyd} suggested as a new approach to maximize this joint posterior instead, in order to jointly constrain the CMB and lensing, and avoiding the calculation of $\hat g^{\rm MF}$. 
The joint posterior, `$p^{\rm J}$', can be obtained by requesting that $\Xdat - \B \Da \Xunl$ must be the noise: up to irrelevant constants
\begin{align}\label{eq:Jpost}
-2\ln	p^{\rm J}(\Xunl \valpha|\Xdat ) = (\Xdat - \B \Da \Xunl)\cdot N^{-1}(\Xdat - \B \Da \Xunl)+ \Xunl \cdot \:C^{X,-1} \Xunl  + \kappa \cdot \:C^{\kappa, -1} \kappa
\end{align}
The last two terms are the priors on the unlensed CMB and lensing fields.
The posterior is always quadratic in $\Xunl$ at fixed lenses, with a well-defined maximum point. A short calculation confirms that this maximum point conditioned on $\field$ is the exact same Wiener-filtered CMB as produced by the original MAP method just written above. 

For this reason, we can write the joint maximization of \eqref{eq:Jpost} as a maximization problem for $\valpha$ only, after replacing $\Xunl$ by the Wiener-filtered CMB in the posterior function. Again, discarding irrelevant constants, a short calculation shows that this function to the be maximized is then perfectly equivalent to the quadratic piece of the CMB-marginalized likelihood and prior

\begin{align}
	&\left.-2\ln	p^{\rm J}(\Xunl\valpha|\Xdat )\right|_{\Xunl = \XWF} =\Xdat \cdot \Covai\Xdat  + \kappa \cdot \:C^{\kappa, -1} \kappa + \text{   const}\nonumber
\end{align}
This is identical to the lensing-only posterior~\eqref{eq:pm}, but without the determinant term. Since it is the determinant term that sources the mean-field, the joint maximization problem is thus perfectly identical to that of a marginalized MAP reconstruction that completely discards all sources of mean-fields.  Direct calculation of the gradients of \eqref{eq:Jpost} can also be used to confirm this. The equation for $\Xunl$ sets it equal to $ \XWF$, and that for $\valpha$ gives then $\hat g^{\rm QD} + \hat g^{\rm Pri} = 0$, which is the marginalized MAP equation, setting the mean-fields to zero.

Hence, both maximization methods reconstruct a lensing map and partially-delensed CMB fields, and, if implemented consistently, they must give identical results in the case that the `marginalized' method neglects completely the mean-field contamination at each step of the iterative search.

\section{Beyond-QE response and biases}\label{app:biases}
\renewcommand{\Amat}[2]{\ensuremath{\mathsf{A}_{#1#2}}}
We can make explicit connection to standard QE power spectrum terms perturbatively around the delenser $\kappa$ using
\begin{equation}
	\bar C^d_{\field} -  \bar C^s_{\field} \sim \sum_{L'M'} \left(\bar C^s_{\field} f^{\field \dagger}_{L'M'} \bar C^s_{\field}\right) \:\kappa^{\rm res}_{L'M'} + \cdots
\end{equation}
where $\kappa^{\rm res} = \kappa^{\rm True} - \kappa$.
With this, we have
\begin{itemize}
\item Linear response of $\gQD$ to residual lensing $\kappa^{\rm res}_{L'M' }$:
\begin{align}
	\mathcal R^\kappa_{LM, L'M'}& = \frac 12 \Tr \left[f_{LM}^\kappa \bar C_{\valpha}^s f_{L'M'}^{\kappa \dagger} \bar C_{\valpha }^{s}\right] \\
\end{align}
\item Primary term \eqref{eq:1st}:
\begin{equation}
\av{\left|\hat g_{LM}\right|^2}  \ni \sum_{L'M'} C^{\text{res}}_{L'}\left| \mathcal R^\kappa_{LM, L'M'}\right|^2
\end{equation}
\item Gaussian $n^{(0)}$-bias (the one present in the absence of any residual lensing).
\begin{align}
	\av{\hat g_{LM}\hat g^*_{L'M'}} \ni n^{(0)}_{LM, L'M'} 
	=\frac 12 \Tr \left[f_{LM}^\kappa \bar C_{\valpha}^s f_{L'M'}^{\kappa \dagger} \bar C_{\valpha }^{s}\right] = \mathcal R^{\kappa}_{LM, L'M'} 
\end{align}
\item $n^{(1)}$-like contribution to the spectrum linear response to the residual lensing spectrum
\begin{align}
 \av{\hat g_{LM}\hat g^*_{L'M'}} \ni n^{(1)}_{LM, L'M'}
	=\sum_{\vL \vM}C_{\vL}^{\text{res}} \frac 12 \Tr \left[f^{\kappa}_{LM}\bar C_{\valpha}^sf^{\kappa \dagger}_{\vL \vM}\bar C_{\valpha}^sf^{\kappa\dagger}_{L'M'}\bar C_{\valpha}^sf^{\kappa}_{\vL \vM}\bar C_{\valpha}^s \right] 
\end{align}
\item Mean-field:
\begin{equation}
	\av{\hat g^{\text{QD}}_{LM}} = \frac 12 \Tr \left[f^\field_{LM} \bar C_\field^s\right]
\end{equation}
\end{itemize}
All terms reduce to their standard and isotropic QE GMV counterparts for vanishing delenser $\kappa$ and an otherwise isotropic configuration.
\section{Gradients are quadratic estimators}\label{app:gradient}
\newcommand{\bxdel}[0]{\bar X^{\rm del}}
\newcommand{\Beam}[0]{\B}
Here we discuss the precise relation between the CMB lensing likelihood gradients (`$\hat g$' )and quadratic estimators on delensed maps (`$\hat q$'). Here we start with the explicit expression of the CMB lensing likelihood gradient quadratic piece, which we formally manipulate to make this structure clearer. As given in the main text, the prefactor between $\hat g$ and $\hat q$ corresponds to the Jacobian between the reparametrized fields, providing the simple physical justification for this.

As shown in~\cite{Belkner:2023duz}, the quadratic piece of the gradient with respect to the deflection vector field components is, to a very good approximation,
\begin{equation}\label{eq:gqd}
	\hat g^{\rm QD}(\hn) = -\bar X(\hn) \left[ \Da \eth \hat X^{\rm WF}\right](\hn)
\end{equation}
This discards a term sourced by the sky curvature, which is second order in the deflection angle and can be safely neglected for realistic CMB lensing deflection fields.
In this equation, the second map on the right is the remapped gradient of the Wiener-filtered CMB $\hat X^{\rm WF}$. The Wiener-filtered CMB is the most probable CMB conditioned on the likelihood model, also given in appendix~\ref{MvsJ}. Two equivalent expressions for this map in harmonic space are
\begin{align} \label{eq:XWF}
	\hat X^{\rm WF} &=  (C^{X,-1} + \Nai)^{-1} \Da^{\dagger}\Beam^\dagger N^{-1} X^{\rm dat} = C^X \Da^\dagger \Beam^\dagger \Covai X^{\rm dat},
\end{align}	
The harmonic space inverse noise matrix in that equation is defined through
\begin{equation}
\Nai \equiv \Da^\dagger \Beam^\dagger N^{-1} \Beam \Da.
\end{equation} 
On Eq.~\eqref{eq:gqd}, the map $\bar X$ is the inverse-variance filtered leg, with equivalent expressions
\begin{align}
	\bar X &= \Beam^\dagger N^{-1} \left(X^{\rm dat} - \Beam \Da \hat X^{\rm WF} \right)= \Beam^\dagger \Covai X^{\rm dat} = \mathcal B^\dagger \bX. \label{eq:ones}
\end{align}
We now make the assumption that the deflection field is invertible. This assumption is not needed for the lensing reconstruction, but is harmless in practice provided other sources of anisotropies present in the data are properly accounted for. 
Then, deflection by the inverse is represented by the inverse operator $\Da^{-1}$. It also follows that $\Da^{-\dagger} = |A_{\field}| \Da$. One can show from this that the unit matrix in position space can be written as
\begin{align}
\delta^D(\hn_1, \hn_2) &= \left[\Da^{-\dagger} \Da^{\dagger}\right](\hn_1, \hn_2)  = |A_{\field}|(\hn_1) \left[\Da \Da^\dagger\right](\hn_1, \hn_2). 
\end{align}
The key point is to insert this big unit matrix in front of $\bar X$, obtaining this representation,
\begin{equation}\label{eq:MF} 
		\hat g_{\field}^{\rm QD}(\hn) = |A_{\field}|(\hn)\left[ \Da \hat q \right](\hn), \quad\text{ with }\quad
	\hat q(\hn) \equiv  - \left[ \Da^{\dagger}\bar X\right](\hn)\: \eth \hat X^{\rm WF}(\hn)
\end{equation}
The claim is that $\hat q$ as defined there can be seen as a standard GMV quadratic estimator acting on delensed maps. 

To see this, make the assumption that the matrix $\Nai$ can be inverted, with an inverse that we write $\Na$. This assumption does not hold and is not useful for lensing reconstruction in practice anyways. We only use it as a notational trick in the following lines.  We can then define a beam-deconvolved and delensed CMB as
\begin{equation}
		 X^{\rm del} \equiv \Na \Da^{\dagger }\Beam^\dagger N^{-1} X^{\rm dat}.
\end{equation}

In the case that the data beam match that of the fiducial model, and that the true deflection is $\field$, the signal part of $X^{\rm del}$ is the unlensed CMB. Under the same assumptions, the covariance of $X^{\rm del}$ is simply
\begin{equation}
	 \av{X^{\rm del}X^{\dagger, \rm del}} = C^X + \Na.
\end{equation}
Using a bit of matrix algebra, the quadratic piece can then be written
\begin{equation}\label{eq:gQE_del}
		\hat q(\hn) = -\bxdel(\hn)\: \eth X^{\rm WF, \rm del}(\hn)
\end{equation}
with the following definitions:
\begin{align}
	 X^{\rm del, \rm WF} &\equiv  C^X \left[ C^X + \Na \right]^{-1}\: X^{\rm del} \quad\text{and}\quad
	 \bxdel \equiv   \left[ C^X + \Na \right]^{-1}\: X^{\rm del}
\end{align}
Written in this way, $\hat q(\hn)$ displays explicitly the exact same form of standard QE applied to $X^{\rm del}$, using the anisotropic noise model in the filters. While we made the generally incorrect assumption of the invertibility of $\Nai$ to write $\hat q$ in this very transparent way, $\hat q$ can very easily be produced in practice without that assumption, since 
\begin{equation}
	 X^{\rm WF, del} = X^{\rm WF}\quad \text{and} \quad	\bxdel = \Da^\dagger \bar X = C^{X,-1} \hat X^{\rm WF}
\end{equation} always hold. 
\section{More details on lensing curvature}\label{app:curvature}
In this appendix we provide some details on the realization-dependent likelihood curvature matrix,
\begin{equation}
	H_{LM, L'M'} \equiv - \frac{\delta^2 \ln p(X|\field)}{\delta \field_{LM} \delta \field^*_{L'M'}}.
\end{equation}
This matrix was derived fully, in flat-sky notation, in appendix A of \cite{Carron:2018lcr}, though the connection to QE and realization-dependent debiasing was not made there. This connection can be performed, for example, realizing that by definition $ \Covai = \av{\bar S_{\field}\bar S^\dagger_{\field}}$ and manipulating.
This leads to the representation
\begin{align}
H_{LM L'M'} =	&\av{ 4\:  \hat g_{LM}\left[ \bXdat{\kappa}, \bXsim \right]\hat g^{*}_{L'M'}\left[ \bXdat{\kappa}, \bXsim \right] -2\:\hat g_{LM}\left[ \bXsim, \bXsimp \right]\hat g^*_{L'M'}\left[ \bXsim, \bXsimp \right]}_{\Xsim \Xsimp|\field} \\&- \frac 12 \bXdat{\field}\cdot f^{(2)\field}_{LML'M'} \bXdat{\field} + \av{\frac 12 \bXsim \cdot f^{(2)\field}_{LML'M'} \bXsim}_{S|\field}
\end{align}
In the first line, only the quadratic part of the gradient is relevant, since the mean-field would anyway not contribute. In the main text, we defined the $M$-averaged diagonal of the first line the realization-dependent debiaser RD-$\hat n_L^{(0)}$. This matrix contains other relevant realization-dependent effects, discussed in~\ref{app:GI}. The $M$-averaged diagonal of the second line probes delensed 2-pt information, as we discuss in~\ref{app:del2pt}.
It is often useful to consider the curvature with respect to the 2 components of the deflection vector field $\nabla_a \phi$, instead of just $\kappa$. In this case the matrix takes on 2 additional indices, but has often simpler position-space representation. 
For invertible deflection fields, the matrix is also often simpler for the remapped fields, as defined in Eq.~\eqref{eq:gvsq} (the `delensed delenser'). This is because it is more isotropic in this space -- the signal has been delensed and remapping effects only acts onto the noise matrix. This corresponds to the operation
\begin{equation}\label{eq:H2tildeH}
	H = \Da^{-\dagger} \widetilde H \Da^{-1}
\end{equation}
where $\widetilde H$ is this `more isotropic' curvature defined with respect to the delensed delensers~\eqref{eq:gvsq}.

\subsection{Gradient inversion}\label{app:GI}
Consider ignoring mean-fields, which is often adequate on small scales. One operational consequence is to remove in the curvature matrix all terms without explicit dependence on the data. Another consequence is that $\bar X$ is driven to zero close to a maximum likelihood point: the mean-field free gradient is the quadratic piece, itself proportional to $\bar X$ at each point, Eq.~\eqref{eq:gqd}).
In these conditions, the only relevant term in the likelihood curvature matrix close to a maximum likelihood point is (in flat-sky notation)
\begin{equation}
 H_{ab}(\hn_1, \hn_2) \sim \nabla_a	\hat X^{\rm WF}(\hn'_1) K(\hn_1,\hn_2)\nabla_b\hat X^{\rm WF}(\hn_2') \quad \text{where}\quad 	K \equiv \Beam^\dagger \Covai \Beam.
\end{equation}
As usual, $\hat n'$ is the undeflected position corresponding to $\hat n$ according to the delenser, $\hat n' \sim \hat n + \nabla \phi(\hn)$.
Viewed as a matrix acting on arbitrary lensing deflection vector fields, $H$ sends to zero any vector field orthogonal to the undeflected, Wiener-filtered, CMB gradient at every point. This is expected since those modes have no leading order map-level effects to the CMB and can only be poorly constrained.

However, the CMB lensing likelihood gradient~\eqref{eq:gqd} is parallel at each point to this CMB gradient by construction. In this subspace of gradient-parallel vector fields a (formal) matrix inverse can be found,
\begin{equation}
	H^{-1}_{ab}(\hn_1, \hn_2) =\frac{\nabla_a	\hat X^{\rm WF}(\hn_1')}{|\nabla \hat{X}^{\rm WF}(\hat n_1')|^2} K^{-1}(\hn_1, \hn_2) \frac{1}{\nabla_b \hat X^{\rm WF}(\hn_2')}.
\end{equation}
Applying this to $\hat g$  to normalize it gives a formal gradient inversion expression instead of a quadratic estimator,

\begin{equation}
	\nabla_a \hat {\phi}(\hat n)\sim  \frac{\nabla_a	\hat X^{\rm WF}(\hn')}{|\nabla \hat{X}^{\rm WF}(\hat n')|^2}\widetilde X(\hn)
\end{equation}

where $\widetilde X =  \Beam^{-1} X$ is the formal beam-deconvolved data map.
\subsection{Delensed 2-pt information}\label{app:del2pt}
Part of the lensing spectrum score function is the $M$-average of the diagonal of the non-perturbative second order response terms. We justify the intuitive claim that this is (close to) equivalent to the 2-pt information of the delensed CMB, and hence carries little independent information.
We consider spin-0 temperature for simplicity. We need the log-likelihood second variation $-\frac 12\bX \cdot f^{(2)\kappa} \bX - \av{\cdots }_{S|\field}$ at non-zero delenser with respect to the spin-weighted deflection field components. In order to compute variations, we can use the fact that to a very good approximation (neglecting sky curvature on the scale of the deflection), a variation with respect to the deflection field is the same as i) taking a gradient, and ii) remap it with the delenser~\cite[Eq.~18]{Belkner:2023duz}. Doing so twice on the log-likelihood gives
\begin{equation}
H_{ab}(\hn_1, \hn_2) \ni \bXdat{}(\hn_1) \left[\frac 14 \sum_{\ell m} C^X_l \:  (\eth^{-a} Y_{\ell m})(\hn'_1) (\eth^{-b}Y_{\ell m})^*(\hn'_2) \right]\bXdat{}(\hn_2) - \frac 14\delta^{D}(\hn_1,\hn_2) \bXdat{}(\hn_1)  \eth^{-a}\eth^b\hat{X}^{\rm WF}(\hn_2').
\end{equation}
We give here only the data-dependent terms. $\bXdat{}$ is $\mathcal B^\dagger \bX$, Eq.~\eqref{eq:ones}.
All explicit anisotropies disappear when considering the curvature with respect to the delensed delenser. We can again do that by inserting a factor $\Da^{-\dagger} \Da^{\dagger} $ in front of $\bXdat{}$, and using~\eqref{eq:H2tildeH},
\begin{equation}
	\widetilde H_{ab}(\hn_1, \hn_2) \ni \bxdel(\hn_1)\left[\frac 14 \sum_{\ell m} C^X_l \:  (\eth^{-a} Y_{\ell m})(\hn_1) (\eth^{-b}Y_{\ell m})^*(\hn_2) \right]\bxdel(\hn_2) - \frac 14\delta^{D}(\hn_1,\hn_2) \bxdel(\hn_1)  \eth^{-a}\eth^b\hat{X}^{\rm WF}(\hn_2).
\end{equation}
Consider the first term. Projecting the matrix onto harmonic space by integrating over $\hn_1$ and $\hn_2$, with harmonic indices $LM$ and $L'M'$, and expanding both $\bxdel$ in harmonics ($\ell_1 m_1, \ell_2 m_2$), produces a pair of Gaunt integrals, and hence a pair of Wigner 3j symbols. There is already one common $m$ sum connecting the pair of 3j symbols. $M$-averaging the diagonal ($L=L'$, $M=M'$) leads to a second common magnetic moment summation, causing the entire 3j product to collapse ($\ell_1 = \ell_2, m_1 = m_2$). The second term is more straightforward. The result is then simply that the second order response terms of
\begin{equation}
 \sum_{M=-L}^L \widetilde H_{LMLM} \text{ are a linear combination of the pseudo power of $\bxdel$ (or equivalently of that of $\hat X^{\rm WF}$)}
\end{equation}
Hence as a statistics, $\sum_{M}\widetilde H_{LMLM}$ is completely degenerate to the delensed 2-pt power, carrying no independent information.
The lensing spectrum score is the $M$-averaged diagonal of $H$, not $\widetilde H$, hence the degeneracy is not expected to be perfect owing to the projection from delensed to observed space, but is still fully there to leading order.

\end{widetext}

\end{document}